\documentclass[twocolumn,prl,superscriptaddress,showpacs,aps,10pt]{revtex4-2}
\newcommand{\pagenumbaa}{1}
\usepackage{graphicx}
\usepackage{color}
\usepackage{float}
\usepackage[normalem]{ulem}

\usepackage{verbatim}		

\usepackage[colorlinks=true,linkcolor=blue,citecolor=blue,urlcolor=blue]{hyperref}
\usepackage{units}
\usepackage{units}
\usepackage{braket}

\begin{document}


\title{High mobility SiMOSFETs fabricated in a full 300mm CMOS process}


\author{Timothy N. Camenzind}\altaffiliation{These authors contributed equally to this work}
\affiliation{Department of Physics, University of Basel, Klingelbergstrasse 82, 4056 Basel, Switzerland}
\author{Asser Elsayed}\altaffiliation{These authors contributed equally to this work}
\affiliation{Department of Physics and Astronomy, KU Leuven, Celestijnenlaan 200D, B-3001 Leuven, Belgium}
\affiliation{IMEC, Kapeldreef 75, B-3001 Leuven, Belgium}
\author{Fahd A. Mohiyaddin}
\affiliation{IMEC, Kapeldreef 75, B-3001 Leuven, Belgium}
\author{Ruoyu Li}
\affiliation{IMEC, Kapeldreef 75, B-3001 Leuven, Belgium}
\author{Stefan Kubicek}
\affiliation{IMEC, Kapeldreef 75, B-3001 Leuven, Belgium}
\author{Julien Jussot}
\affiliation{IMEC, Kapeldreef 75, B-3001 Leuven, Belgium}
\author{Pol Van Dorpe}
\affiliation{Department of Physics and Astronomy, KU Leuven, Celestijnenlaan 200D, B-3001 Leuven, Belgium}
\affiliation{IMEC, Kapeldreef 75, B-3001 Leuven, Belgium}
\author{Bogdan Govoreanu}
\affiliation{IMEC, Kapeldreef 75, B-3001 Leuven, Belgium}
\author{Iuliana Radu}
\affiliation{IMEC, Kapeldreef 75, B-3001 Leuven, Belgium}
\author{Dominik M. Zumb\"uhl}
\affiliation{Department of Physics, University of Basel, Klingelbergstrasse 82, 4056 Basel, Switzerland}


\begin{abstract}
The quality of the semiconductor-barrier interface plays a pivotal role in the demonstration of high quality reproducible quantum dots for quantum information processing. In this work, we have measured SiMOSFET Hall bars on undoped Si substrates in order to investigate the quality of the devices fabricated in a full CMOS process. We report a record mobility of \unit[$17.5\times 10^{3}$]{cm$^2$/Vs} with a sub-\unit[10]{nm} oxide thickness indicating a high quality interface, suitable for future qubit applications. We also study the influence of gate materials on the mobilities and discuss the underlying mechanisms, giving insight into further material optimization for large scale quantum processors.
\end{abstract}

\maketitle

\setcounter{page}{\pagenumbaa}
\thispagestyle{plain}


The spin of an electron in Silicon has been considered as one of the most promising candidates for large-scale quantum computers, due to its long coherence time, compactness, potential to operate at relatively high temperatures, and compatibility with CMOS technology for upscaling \cite{Vandersypen2017interfacing, Petta2005coherent, Veldhorst2014addressable}. High fidelity single and two qubit operations have been demonstrated in academic lab-based devices \cite{Maurand_2016, Veldhorst_2015}. The scale up process to multi-qubit arrays will however entail high quality qubits in large numbers, necessitating a transition to industrial fabrication techniques \cite{Li_2018}.

While spin qubit-devices can be fabricated using semiconductor manufacturing techniques, a detailed and careful analysis on the impact of different fabrication process steps on the performance of spin qubits is crucial. Hallbar mobility is a widely used metric to characterize the MOS interface quality, which can provide valuable feedback to fabrication process optimization. In this work, we perform transport measurements on SiMOSFETs and investigate the quantum Hall effect. The gate stack for SiMOSFET is the same as that for qubit integration \cite{Li_2020}. We extract key characteristics of the MOS gate stack such as critical density and mobility as well as transport and quantum lifetimes. Comparing the obtained results to a transport model, we identify and quantify the leading scattering terms. Additionally, we explore possible mobility limiting factors when changing the top gate material from PolySi to titanium nitride (TiN).

The devices measured in this work are undoped inversion gated Hall bars fabricated in a state-of-the art 300 mm fab line \cite{Li_2020}. The starting substrate consists of a (100) silicon wafer with a background doping \textless  \unit[$5\times 10^{15}$]{cm$^{-3}$}. The Hall bar top gates are defined by electron beam lithography and subsequent dry etching process, and the fanout metal pins are defined by deep UV lithography. After depositing the gate material, the whole wafer is covered with a passivation layer. In order to measure the electronic properties of the investigated samples, a positive voltage has been applied to the top gate to form a 2-dimensional electron gas (2DEG) at the interface of Si and SiO$_2$. Throughout this work, we perform four-wire electronic transport measurements using standard lock-in techniques. Both the longitudinal and transverse Hall voltages were measured simultaneously with differential voltage amplifiers. Additionally, the current through the sample was measured using an IV transimpeadance amplifier. The measurements were carried out in a dilution refrigerator with a base temperature of \unit[20]{mK} with a \unit[9]{T} magnet. Higher temperatures were achieved by heating the mixing chamber, resulting in electron temperatures between \unit[100]{mK} and \unit[4.2]{K}.

\begin{figure}
\centering
\includegraphics[scale=1]{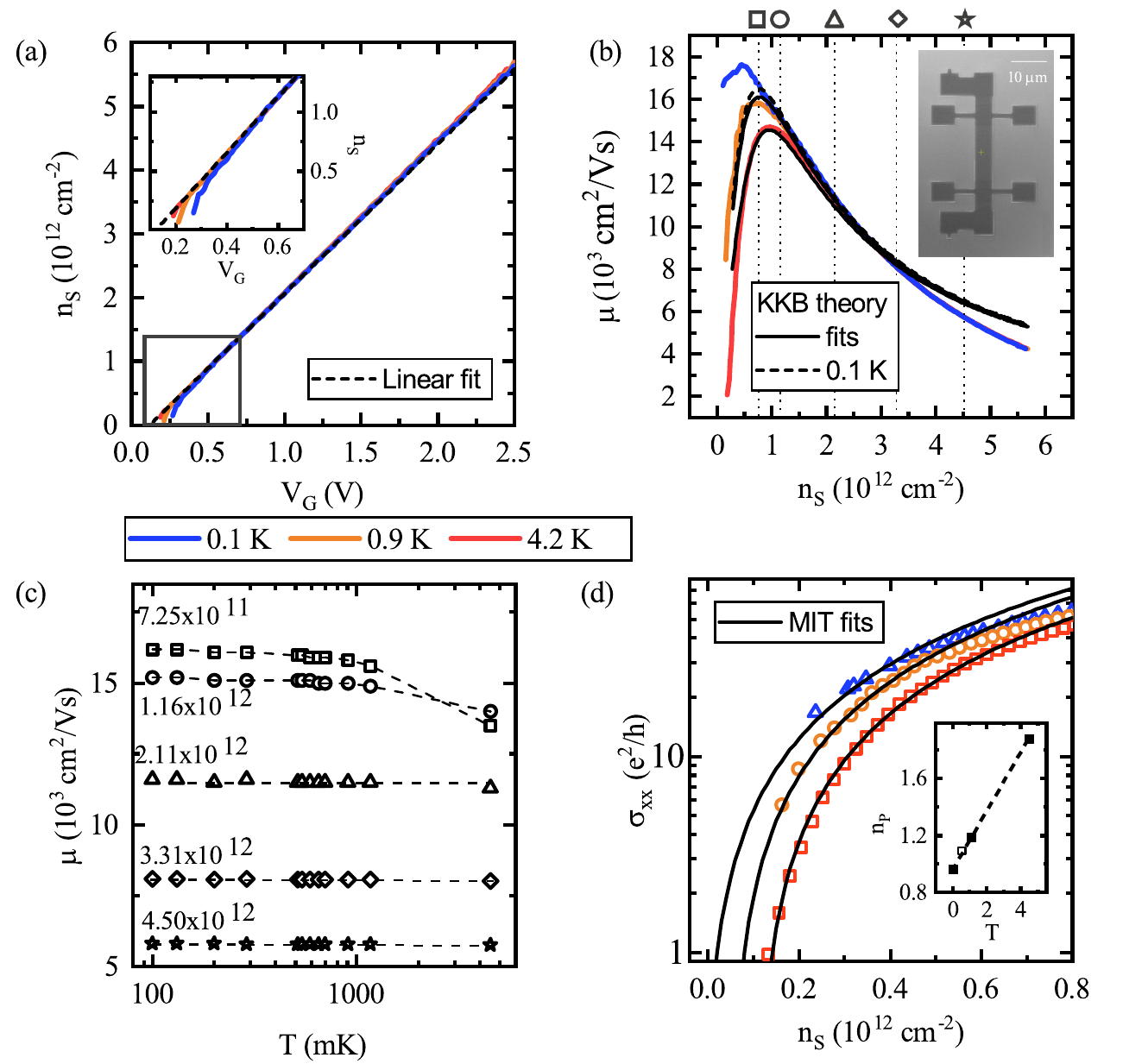}
\caption{\label{F_1}\textbf{Density and Mobility study} (a) Linear relationship between 2DEG sheet density $n_S$ and the top gate voltage $V_{G}$. Below a top gate voltage of $\approx$ \unit[0.4]{V} the extracted density deviates from the expected linear behaviour as shown in the inset. (b) Channel mobility $\mu$ measured as a function of $n_S$ and corresponding fit to a mobility model (black) which includes scattering from charge impurities and surface roughness. The inset shows a SEM image of a similar device as the one used during the experiments with the same dimensions. (c) Channel mobility $\mu$ trends with respect to temperature at various sheet densities. The symbols coincide with the ones used in panel b. The dashed lines are linear fits to the extracted mobilities and appear curved due to the logarithmic scale. (d) Longitudinal conductivity $\sigma_{xx}$ in the low density range and fit to a percolation theory (black). A linear trend with respect to temperature for $n_p$ is observed in the inset.}
\end{figure}

Fig.\ref{F_1} shows the density and mobility study performed on a sample with a PolySi top gate at three different temperatures. First, the density of charge carriers in the 2DEG was extracted from the transverse Hall effect in the linear regime far from quantum oscillations, particularly at the lowest temperature with magnetic field between $\pm$\unit[1]{T}. The density of charge carriers in the 2DEG has an overall linear dependence on the applied top gate voltage $V_{G}$ as expected from a plate-capacitor model. Assuming the relative dielectric constant $\epsilon _r (\mathrm{SiO_2})=3.9$, we extract the thickness of the oxide and obtain a value of \unit[ 8.1 $\pm$ 0.1]{nm}. This in good agreement with the oxide thickness of \unit[8]{nm} defined during the fabrication process. As one can see in the inset of Fig.\ref{F_1}a, the extracted density deviates from a linear regime at very low $V_{G}$ near the threshold voltage, particularly at the lowest temperature. A temperature dependent dielectric constant would result in a temperature dependent slope of the $n_S$ regardless of the $V_{G}$ \cite{Lautenschlager_1987}, which is not observed in this study. Partial localization of charge carriers in the disorder potential at low sheet density, on the other hand, could explain this \cite{Tsukada_1976}. Here, the disorder potential is mainly due to bulk charge scattering and scattering from oxide charges. At these rather low densities close to the 2D Metal to Insulator Transition (MIT), partial localization would effectively reduce the sheet density of mobile carriers below a simple capacitor model, as observed, and thus slightly enhance the mobility, which is inversely proportional to the density.

The extracted densities, in combination with the sheet conductance from the longitudinal voltage measurements, allow for the calculation of the mobility $\mu$ at various temperatures. The resulting curves are shown in Fig.\ref{F_1}b. We measured a maximum mobility of \unit[$17.5\times 10^{3}$]{cm$^2$/Vs}, among the highest mobility reported for SiMOSFET devices with sub \unit[10]{nm} oxide thickness \cite{Shankar_2010,Rochette_2019,Sabbagh_2019}. We fit the mobility curve with the Kruithof-Klapwijk-Bakker (KKB) model (Ref.~\cite{Kruithof_1991}) which includes remote impurity scattering and surface roughness terms \cite{Kruithof_1991,Gold_1986}. Solving the model self-consistently for three separate temperatures (0.5, 0.9 and \unit[4.2]{K}) yields the charged impurity center density $N_C= $\unit[ $1.53\pm 0.03\times 10^{10}$]{cm$^{-2}$}, the rms height of the amplitude of the interface roughness $\Delta= $\unit[ 5.9 $\pm$ 0.51]{\AA} and the correlation length $L= $\unit[ 1.8 $\pm$ 0.35]{nm}. These numbers indicate a smooth and clean interface which is critical for future qubit implementations.

As shown in Fig.\ref{F_1}b, there is a slight deviation of the maximum mobility from the theory curve at \unit[0.9]{K}, and the theoretical curve does not agree with the reported data for \unit[0.1]{K} in the low density regime. This could be explained by the assumption of a linear relationship between mobility and temperature in the KKB theory. According to our fitting, this assumption seems to be only valid above \unit[1]{K} and can become a more complex function of temperature below \unit[1]{K} if the low density regime becomes non-linear. This is due to the fact that the slightly reduced density (compared to the capacitor model) results in a slightly higher mobility, explaining the deviation of our data from the theory. Furthermore, for sheet densities $n_S$ close to the depletion charge density $n_{dep}$, the calculation of the finite extent of the wave function in the KKB theory can become unreliable because of the assumed approximations. The theoretically predicted curve can be recovered by using a linearly extrapolated density during the calculation of the mobility. Further details are discussed in Appendix A.

Another key parameter is the percolation density, which indicates the minimum density required to form a conducting channel. Generally, the percolation density describes the disorder in the channel at low densities, where quantum devices typically operate. In Fig.\ref{F_1}d, we employed a MIT model in order to fit the density-dependent conductivity $\sigma _{xx} \sim (n - n_p)^p$ \cite{Tracy_2009}, where $n_p$ and $p$ are the percolation density and universal exponent, respectively. However, this model is only valid in the low density regime and deviates above $n=$\unit[$ 8 \times 10^{11}$]{cm$^{-2}$}. We fixed the exponent $p=1.31$, as expected in a two-dimensional system, and allowed $n_p$ to vary \cite{Tracy_2009}. Our fits yield $n_p = $\unit[ $1.86 \times 10^{11}$]{cm$^{-2}$} at $T= $\unit[ 4.2]{K}. As seen in the inset of \ref{F_1}d the extracted $n_p$ decreases with decreasing temperature, which is in agreement with previous studies \cite{Tracy_2009}, and is below \unit[$1 \times 10^{11}$]{cm$^{-2}$} at the lowest temperature. Such a low percolation density is among the lowest reported for similar devices\cite{Sabbagh_2019, Kim_2017}. The small percolation density and high peak mobility indicate low interface disorder of the MOS gate stack, which is crucial for large scale quantum dot arrays.

Additionally, we investigate the temperature dependence of mobility at various top gate voltages. For sheet densities less than \unit[$2 \times 10^{12}$]{cm$^{-2}$} the mobility is a function of temperature, with a linear dependence for temperatures above \unit[1]{K} (see Fig.\ref{F_1}c). The temperature dependence of the mobility is only visible for low electron densities because mobilities at higher densities are limited by surface roughness, which is not temperature dependent. Additionally, the peak mobility shifts to higher densities at higher temperatures due to the activation of background charges limiting the mobility at high temperatures.

\begin{figure}
\centering
\includegraphics[scale=1]{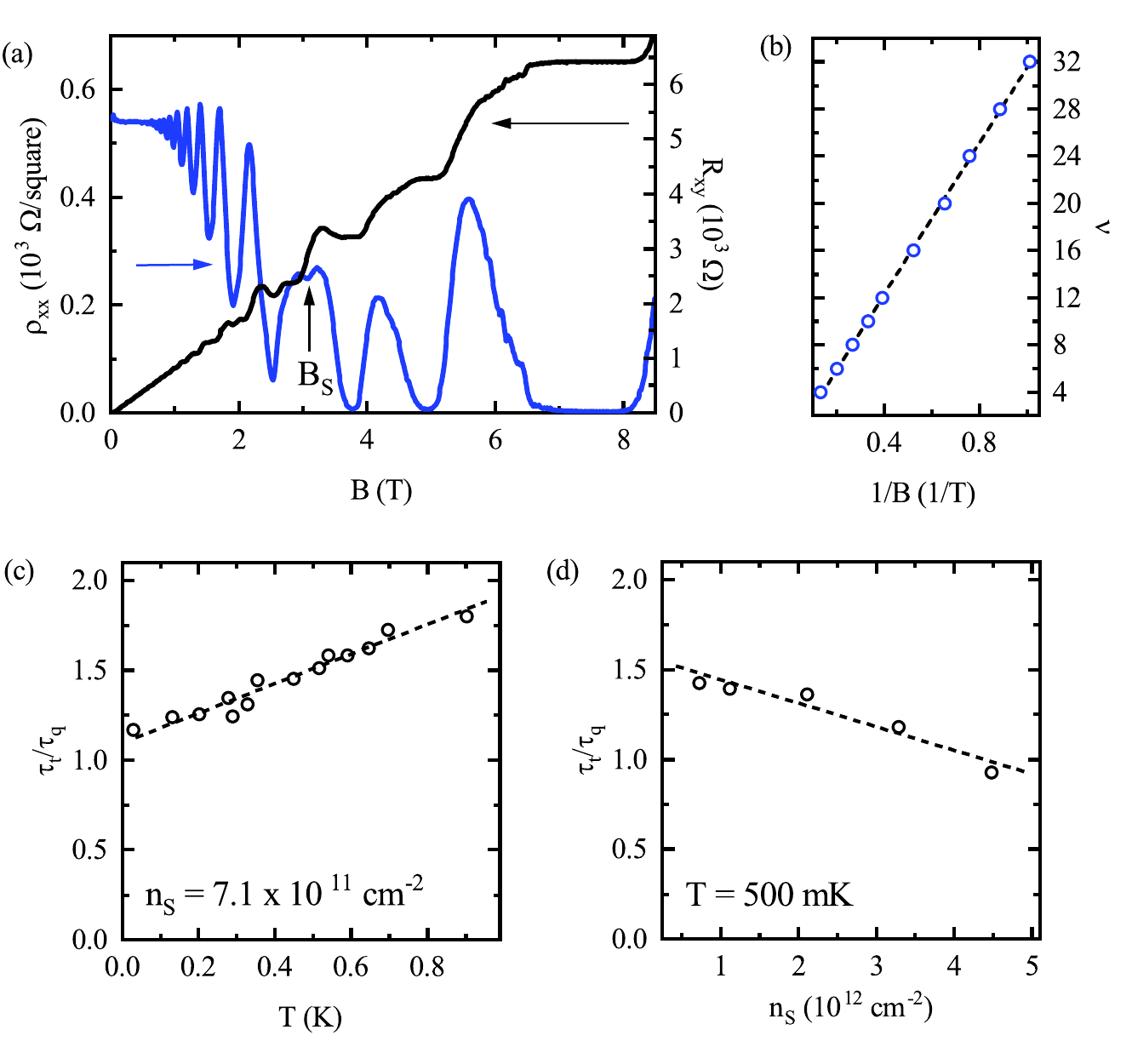}
\caption{\label{F_2}\textbf{Extended Hall studies with transport properties} (a) Longitudinal resistivity $\rho_{xx}$ (blue) and transverse resistance R$_{xy}$ (black) at $n_S = $ \unit[$7.1\times 10^{11}$]{cm$^{-2}$} as a function of magnetic field $B$. The arrow indicates the magnetic field spin is resolved. (b) Linear relationship between filling factor $\nu$ (blue scatter) and inverse magnetic field. The solid line is a fit from which the sheet density is calculated. (c) and (d) Extracted dingle ratio $\tau _t /\tau _q$ as a function of temperature $T$ and sheet density $n_S$, respectively. The dingle ratio is slightly temperature and density dependent and is $\approx 1$, which indicates that large angle scattering is most dominant scattering mechanism in this sample.}
\end{figure}

In order to extract more information about the transport properties and the dominating scattering mechanism, we investigated the transverse resistance R$_{xy}$ and longitudinal resistivity $\rho_{xx}$ in the quantum Hall regime. The magnetic field $B$ was swept up to \unit[8.5]{T} at a fixed sheet density of \unit[$7.1\times 10^{11}$]{cm$^{-2}$} near the peak mobility. As shown in Fig.\ref{F_2}a, we observe Shubnikov-De Haas (SdH) oscillations starting at \unit[0.8]{T} and begin to resolve spin splitting at \unit[3.2]{T}. The $\rho_{xx}$ shows well developed zeros for magnetic fields larger than \unit[3.8]{T}. In combination with the single frequency of the SdH oscillations extracted below \unit[2.5]{T} and the absence of beatings strongly indicate single subband occupancy. We further observe a four-fold degeneracy in the filling factor which changes to a two-fold degeneracy as we increased the magnetic field  (Fig.\ref{F_2}b). The remaining two-fold degeneracy is due to the two lowest lying valleys which are split off from the remaining four valleys due to the z-confinement in Silicon (100) \cite{Herman_1955}. A linear fit of the filling factor  (Fig.\ref{F_2}b)  with respect to inverse magnetic field yields a sheet density \unit[$7.1\times 10^{11}$]{cm$^{-2}$} in agreement with the extracted Hall density, further confirming clean single subband transport in the investigated samples.

Upon further investigations into the SdH oscillation amplitudes below \unit[2.5]{T}, we were able to extract an effective mass $m^*=0.21\pm 0.02\cdot m_0$ from the temperature dependent data, which corresponds to the transverse mass expected for Silicon samples with similar background doping \cite{Ando_1982}. Further details are discussed in Appendix B. From fitting the SdH envelope as a function of temperature and density, we can extract the quantum lifetime $\tau_q$. By combining $\tau _q$, which includes all possible scattering events, and the transport lifetime $\tau _t=e*\mu /m^*$, which includes only the large angle scattering, we extract the so-called Dingle ratio $\tau _t/\tau _q$. The Dingle ratio thus characterizes the dominant scattering mechanism ranging from a ratio of $1$ for predominant large angle scattering to values $>>1$ for dominant small angle scattering. Fig.\ref{F_2}c shows an increase of the Dingle ratio from $\approx 1$ to $\approx 2$ with increasing temperature, indicating a shift in the type of scattering dominating at different temperature ranges. We can see an inverted trend for the Dingle ratio with respect to density. For a fixed temperature of \unit[0.5]{K}, we observe a decrease from a ratio of $\approx 1.5$ to $\approx 1$ by increasing the density, again showing a slight shift towards large angle scattering. This is true for all measured temperatures and densities. In all, the observed Dingle ratio around 1 to 2 shows that short range / large angle scattering is the dominant scattering mechanism in this system, where the scattering centers are near or in the channel itself \cite{Ando_1982}.

\begin{figure}
\centering
\includegraphics[scale=1]{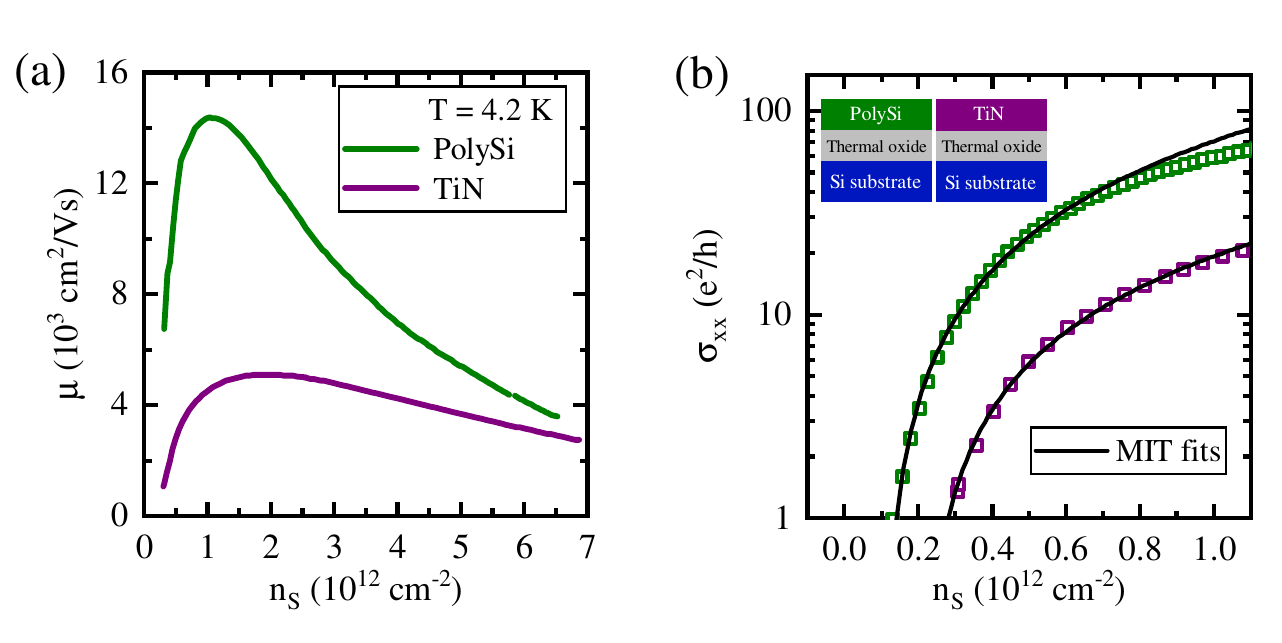}
\caption{\label{F_3}\textbf{Gate material study} (a) The channel mobility $\mu$ with respect to sheet density $n_S$ for PolySi (green) and TiN (violet), both at \unit[4.2]{K}. The mobility shows a decrease by a factor of three when changing the top gate material from PolySi to TiN. (b) Longitudinal conductivity $\sigma_{xx}$ with respect to sheet density. The fits to percolation theory (black) show a significant difference between the PolySi and TiN percolation densities.}
\end{figure}

Furthermore, we investigate the effect of two different gate metals on the mobility and percolation density. We chose PolySi and titanium nitride (TiN) as top gate materials since both are widely applied gating materials for integrated CMOS device fabrication. While PolySi has generally larger mobilities, which indicates a small disorder potential and therefore a low dot-to-dot variation, TiN exhibits low resistivity and is superconducting at low temperatures, which is desirable for microwave qubit control and readout. The density dependent mobilities at \unit[4.2]{K}, for both TiN and PolySi are shown in Fig.\ref{F_3}a. There is a decrease by a factor of three in the maximum mobility when changing the material from PolySi to TiN while all other parameters in the fabrication remain the same. To investigate the underlying mechanism for the mobility difference, we perform high magnetic field Hall measurements on the TiN devices as well. Unlike PolySi which shows vanishing $\rho_{xx}$ in Fig.\ref{F_2}a, $\rho_{xx}$ of TiN does not have well developed zeros. This can be attributed to the lower mobility and could be due to multiple subband transport \cite{Kruithof_1990} (see Appendix C for further details). At high electron concentrations and in the presence of compressive stress, multiple subbands can be occupied and thus introduce inter-subband scattering at low temperatures.

To explain the inter-subband scattering, we firstly consider the effect of strain. On silicon (100) surfaces, materials used as top gates with certain coefficient of thermal expansion (CTE) introduce compressive strain \cite{Nix1941thermal}. This leads to a decrease in the energy splitting of two subsequent subbands \cite{Ando_1982,Kawaji_1976,Eisele_1978}. Strain simulations, described in Appendix D, indicate a large difference in strain between TiN and PolySi. However, the strain could only decrease the TiN mobility by roughly 10 percent rather than the three fold decrease observed \cite{Eisele_1978}. This difference could be explained by the oxygen scavenging of TiN on SiO$_2$ \cite{Filatova_2019}. When depositing TiN on a SiO$_2$ surface, SiO$_2$ gets reduced to SiO$_x$ and the free oxygen combines with TiN to form TiO$_2$ and titanium oxynitride (TiN$_x$O$_y$). This process creates charged oxygen vacancies in the SiO$_2$ layer and because the oxide is very thin in these samples, the charge carriers can then scatter at these defects thus reducing the overall mobility. Indeed, fits of the KKB model to the TiN data indicate a significantly higher $N_C$ than PolySi consistent with the scavenging hypothesis. Those two effects combined therefore could explain the difference in mobilities between the two gate metals.

In summary, we have characterized SiMOSFET samples using Hall effect measurements. We found a record mobility of \unit[$17.5\times 10^{3}$]{cm$^2$/Vs} which is the highest reported mobility to date for SiMOSFET samples with sub-\unit[10]{nm} oxide thickness. Further, we have shown that the dominant mobility limiting factor are charges near the conducting channel, which can be improved by further reducing the impurity levels in the substrate. For example, an epitaxial layer can reach a background doping level of roughly \unit[$1\times 10^{12}$]{cm$^{-3}$}, which is three orders of magnitude lower than the samples measured in this study. Additionally, we show that the choice of top gate material directly affects the mobility and percolation density. Factors such as strain and oxygen scavenging could explain the strong influence on both parameters. This demonstrates that the choice of top gate material is of paramount importance. The MOS interface has been widely identified as the source for charge noise and disorder sites, which limits the qubit performance and upscaling. To further improve the qubit device fabrication, more systematical investigations are needed. The Hallbar mobility and the analysis presented in this study could provide valuable insight into the gate stack over a large device area and direct further fabrication optimizations, which could pave the way for a large scale spin qubit processor integration with a full CMOS process.

\section*{Acknowledgements}

This work was performed as part of imec’s Industrial Affiliation Program (IIAP) on Quantum Computing. Additionally, this work was supported by the Swiss Nanoscience Institute (SNI), the Swiss National Science Foundation (SNF), the National Centers of Competence in Research (NCCR) SPIN and the EU H2020 European Microkelvin Platform (EMP) grant No. 824109.

\section*{Author contributions}
T.N.C., A.E. and D.M.Z performed the experiments, analysed the measurements and wrote the manuscript with input from all authors. B.G. designed the test structures. S.K. and J.J. fabricated the devices. F.A.M. performed the strain simulations. The work was completed with assistance from R.L., P.VD. and I.R..

\section*{Data availability statement}
The data supporting the plots of this paper are available at the Zenodo repository at \url{https://doi.org/10.5281/zenodo.xxxxx}.

\section*{Appendix}

\subsection{A. Mobility as a function of linear density and MIT fitting range}\label{A:lin_mob}

\begin{figure}
\centering
\includegraphics[scale=1]{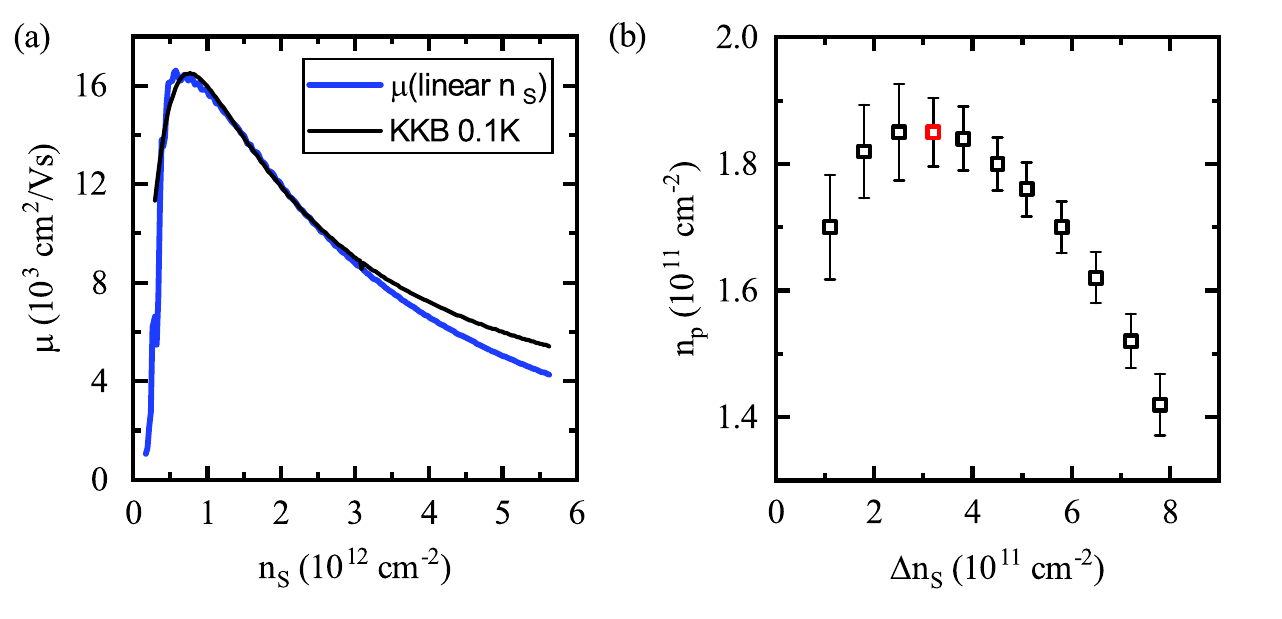}
\caption{\label{S_1}\textbf{Mobility as a function of linear density and MIT fitting range} (a) When using a linearly extrapolated (instead of the carefully extracted) density in order to calculate the mobility of our sample (blue), the peak mobility becomes slightly smaller than previously demonstrated in Fig.\ref{F_1}b. Additionally, we can fit the KKB theory (black) self-consistently with the other temperature data highlighting the importance of the density values. (b) By varying $\Delta n_s$ for extracting the percolation density $n_p$ within the metal to insulator transition (MIT) model, we determine the optimal fitting range. We chose $\Delta n_s=$\unit[$3.2\times 10^{11}$]{cm$^{-2}$} (indicated by the red box) because the extracted $n_p$ does not change significantly while changing $\Delta n_s$.}
\end{figure}

During the calculation of the mobility at the lowest temperature, we observed a peak mobility higher than the KKB model (see Fig.\ref{F_1}b). We associated this deviation to the non-linear density dependence in the small top gate voltage regime at the lowest temperature as shown in the inset of Fig.\ref{F_1}a. This assumption is further supported because the theoretical curve can be recovered using a linear extrapolated density when calculating the mobility (see Fig.\ref{S_1}). This behaviour can be explained within the KKB theory, where the depletion charge density $n_{dep}$ is implicitly assumed to be significantly smaller than the sheet density $n_S$ especially in the critical regime of maximum mobility \cite{Kruithof_1991}. If the density becomes non-linear, this violation of the linear density assumption has a significant impact when calculating the finite extent of the wave function $b$ within the Coulomb scattering term $\braket{|U_C(q)|^2}$ and the dielectric response function $\varepsilon$. Further research in this regime is needed in order to capture the behaviour found in this work, but it exceeds the scope of this work. Nevertheless, we emphasize the mobility of \unit[$17.5\times 10^{3}$]{cm$^2$/Vs} is the properly measured mobility, and the slightly lower peak mobility of \unit[$16.5\times 10^{3}$]{cm$^2$/Vs} in Fig.\ref{S_1} is a modeled mobility assuming a linear density model. The previous mobility record was $\approx$\unit[$10\times 10^{3}$]{cm$^2$/Vs} for a comparable sample although the oxide was slightly thicker in that work\cite{Sabbagh_2019}.

The metal to insulator transition model (MIT) we used in this work is only valid at very low densities (below \unit[$1\times 10^{12}$]{cm$^{-2}$} and therefore is rather sensitive to the fitting range. In order to find the appropriate fitting range, we varied both the minimum and maximum value independently. As shown in Fig.\ref{S_1}, we determined that fitting the data between \unit[$2.7\times 10^{11}$]{cm$^{-2}$} and \unit[$6.0\times 10^{11}$]{cm$^{-2}$} yields the most reliable outcome because variations in the fitting range do not significantly change the result.

\subsection{B. Extraction of effective mass and quantum lifetime}\label{A:extract}

In order validate our evaluation approach and our experiments, we extract the effective mass from the temperature dependence of the SdH oscillations. For that, we subtract a polynominal background from the $\rho _{xx}$ data and use the following formula as described in \cite{Bauer1972low, Celik2011determination}
\begin{equation}
\frac{\Delta\rho _{xx}(T)}{\Delta\rho _{xx}(T_0)}=\frac{T}{T_0}\cdot\frac{\sinh \chi (T)}{\sinh \chi (T_0)}
\end{equation}
with $\chi (T)=2\pi ^2k_BT/\hbar\omega _c$ and $\omega _c=eB/m^*$. Here, $T$ is the temperature, $T_0$ the lowest temperature, $k_B$ is the Boltzmann constant, $\hbar$ the reduced Plank constant, $e$ the electric charge, $B$ the magnetic field and $m^*$ the effective mass. Using the effective mass as the only free fit parameter, we extracted $m^*=0.21\pm 0.02\cdot m_0$, which is close to the expected value of $m^*=0.19\cdot m_e$.

Further, the envelope of the SdH oscillations allows for the extraction of the quantum lifetime $\tau_q$. For that, we use the following formula
\begin{equation}
ln\left( \frac{\Delta\rho _{xx}(B)\sinh\chi}{\chi}\right)=C-\frac{\pi m^*}{eB\tau _q}
\end{equation}
Therefore, by analysing the slope of the magnetic field dependent logarithmic term we can extract $\tau_q$ for all measured densities and temperatures. In combination with the transport lifetime $\tau _t$ we can then further extract the dingle ratio as described in the main text.

\subsection{C. TiN data at high fields}\label{A:TiN}

In comparison to the reported data in Fig.\ref{F_2}a, TiN does not show well developed zeroes in the longitudinal resistivity $\rho _{xx}$ at high magnetic fields at a temperature of \unit[500]{mK} (see Fig.\ref{S_4}), as expected for a much lower mobility. Further, the transversal Hall resistance $R_{xy}$ shows oscillating features instead of the expected Hall plateaus. This could be due to multiple subbands contributing to the overall transport or with the increased density of background charges due to the above mentioned oxygen scavenging.

\subsection{D. Strain simulations}\label{A:Strain}

\begin{figure}
\centering
\includegraphics[scale=1]{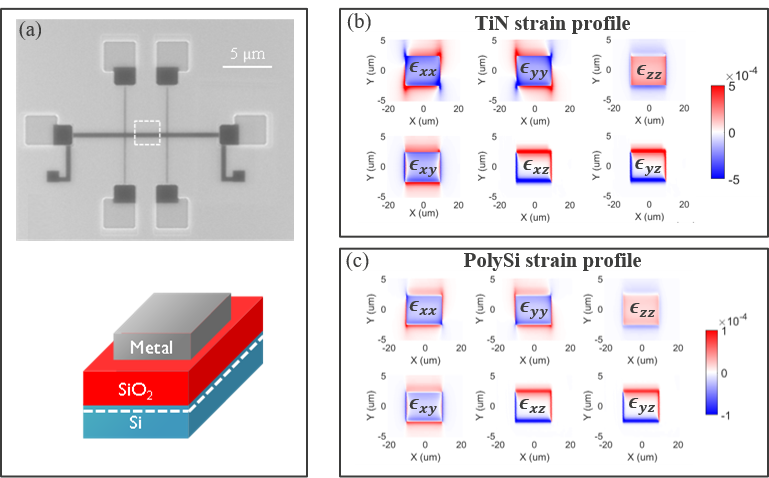}
\caption{\label{S_4}\textbf{Gate material strain simulations} (a) Schematic of the Hall bar used for the strain simulation. The strain in the channel was calculated using sprocess in the region indicated by the dashed white line. (b, c) The strain in the channel for the PolySi and TiN respectively. The model uses literature values for the thermal expansion coefficient for both materials.}
\end{figure}

We quantify the stain induced in the 2DEG due to the different thermal expansion coefficients of both gate materials. The strain was simulated using a commercially available software \cite{Sentaurus}. The silicon was simulated isotropically. The temperature dependence of the CTE for PolySi up to room temperature is assumed to be linear because there is no reported data above \unit[200]{K}\ \cite{Thorbeck_2015}. Further, we neglect the intrinsic strain induced during the fabrication process, which is beyond the scope of this work. Thermal contraction is considered from room temperature down to the cryogenic measurement temperatures. All simulations were performed in three dimensions and the components of strain $\epsilon_x , \epsilon_y , \epsilon_z$ are calculated \unit[5]{nm} below the oxide due to the finite extension of the wavefunction. As shown in Fig.\ref{S_4}b and c, there is a large difference in strain between TiN and PolySi.  While TiN induces strain up to \unit[$2.4\times 10^{-4}$]{}, PolySi induces significantly less strain of approximately only \unit[$0.28\times 10^{-4}$]{}.

\newpage


\begin{thebibliography}{10}

\bibitem{Vandersypen2017interfacing}
L.~Vandersypen, H.~Bluhm, J.~Clarke, A.~Dzurak, R.~Ishihara, A.~Morello,
  D.~Reilly, L.~Schreiber, and M.~Veldhorst.
\newblock \emph{Interfacing spin qubits in quantum dots and donors—hot,
  dense, and coherent}.
\newblock npj Quantum Information \textbf{3}, 1 (2017).

\bibitem{Petta2005coherent}
J.~R. Petta, A.~C. Johnson, J.~M. Taylor, E.~A. Laird, A.~Yacoby, M.~D. Lukin,
  C.~M. Marcus, M.~P. Hanson, and A.~C. Gossard.
\newblock \emph{Coherent manipulation of coupled electron spins in
  semiconductor quantum dots}.
\newblock Science \textbf{309}, 2180 (2005).

\bibitem{Veldhorst2014addressable}
M.~Veldhorst, J.~Hwang, C.~Yang, A.~Leenstra, B.~de~Ronde, J.~Dehollain,
  J.~Muhonen, F.~Hudson, K.~M. Itoh, A.~Morello, \emph{et~al.}
\newblock \emph{An addressable quantum dot qubit with fault-tolerant
  control-fidelity}.
\newblock Nature nanotechnology \textbf{9}, 981 (2014).

\bibitem{Maurand_2016}
R.~Maurand, X.~Jehl, D.~Kotekar-Patil, A.~Corna, H.~Bohuslavskyi,
  R.~Lavi{\'e}ville, L.~Hutin, S.~Barraud, M.~Vinet, M.~Sanquer, \emph{et~al.}
\newblock \emph{A CMOS silicon spin qubit}.
\newblock Nature communications \textbf{7}, 1 (2016).

\bibitem{Veldhorst_2015}
M.~Veldhorst, C.~Yang, J.~Hwang, W.~Huang, J.~Dehollain, J.~Muhonen,
  S.~Simmons, A.~Laucht, F.~Hudson, K.~M. Itoh, \emph{et~al.}
\newblock \emph{A two-qubit logic gate in silicon}.
\newblock Nature \textbf{526}, 410 (2015).

\bibitem{Li_2018}
R.~Li, L.~Petit, D.~P. Franke, J.~P. Dehollain, J.~Helsen, M.~Steudtner, N.~K.
  Thomas, Z.~R. Yoscovits, K.~J. Singh, S.~Wehner, \emph{et~al.}
\newblock \emph{A crossbar network for silicon quantum dot qubits}.
\newblock Science advances \textbf{4}, eaar3960 (2018).

\bibitem{Li_2020}
R.~Li, N.~D. Stuyck, S.~Kubicek, J.~Jussot, B.~Chan, F.~Mohiyaddin, A.~Elsayed,
  M.~Shehata, G.~Simion, C.~Godfrin, \emph{et~al.}
\newblock \emph{A flexible 300 mm integrated Si MOS platform for electron-and
  hole-spin qubits exploration}.
\newblock In \emph{2020 IEEE International Electron Devices Meeting (IEDM)},
  38--3 (IEEE, 2020).

\bibitem{Lautenschlager_1987}
P.~Lautenschlager, M.~Garriga, L.~Vina, and M.~Cardona.
\newblock \emph{Temperature dependence of the dielectric function and interband
  critical points in silicon}.
\newblock Physical Review B \textbf{36}, 4821 (1987).

\bibitem{Tsukada_1976}
M.~Tsukada.
\newblock \emph{On the tail states of the landau subbands in MOS structures
  under strong magnetic field}.
\newblock Journal of the Physical Society of Japan \textbf{41}, 1466 (1976).

\bibitem{Shankar_2010}
S.~Shankar, A.~Tyryshkin, J.~He, and S.~Lyon.
\newblock \emph{Spin relaxation and coherence times for electrons at the Si/SiO
  2 interface}.
\newblock Physical Review B \textbf{82}, 195323 (2010).

\bibitem{Rochette_2019}
S.~Rochette, M.~Rudolph, A.-M. Roy, M.~Curry, G.~T. Eyck, R.~Manginell,
  J.~Wendt, T.~Pluym, S.~Carr, D.~Ward, \emph{et~al.}
\newblock \emph{Quantum dots with split enhancement gate tunnel barrier
  control}.
\newblock Applied Physics Letters \textbf{114}, 083101 (2019).

\bibitem{Sabbagh_2019}
D.~Sabbagh, N.~Thomas, J.~Torres, R.~Pillarisetty, P.~Amin, H.~George,
  K.~Singh, A.~Budrevich, M.~Robinson, D.~Merrill, \emph{et~al.}
\newblock \emph{Quantum Transport Properties of Industrial Si 28/Si O 2 28}.
\newblock Physical Review Applied \textbf{12}, 014013 (2019).

\bibitem{Kruithof_1991}
G.~Kruithof, T.~Klapwijk, and S.~Bakker.
\newblock \emph{Temperature and interface-roughness dependence of the electron
  mobility in high-mobility Si (100) inversion layers below 4.2 K}.
\newblock Physical Review B \textbf{43}, 6642 (1991).

\bibitem{Gold_1986}
A.~Gold and V.~Dolgopolov.
\newblock \emph{Temperature dependence of the conductivity for the
  two-dimensional electron gas: Analytical results for low temperatures}.
\newblock Physical Review B \textbf{33}, 1076 (1986).

\bibitem{Tracy_2009}
L.~Tracy, E.~Hwang, K.~Eng, G.~Ten~Eyck, E.~Nordberg, K.~Childs, M.~Carroll,
  M.~Lilly, and S.~D. Sarma.
\newblock \emph{Observation of percolation-induced two-dimensional
  metal-insulator transition in a Si MOSFET}.
\newblock Physical Review B \textbf{79}, 235307 (2009).

\bibitem{Kim_2017}
J.-S. Kim, A.~M. Tyryshkin, and S.~A. Lyon.
\newblock \emph{Annealing shallow Si/SiO2 interface traps in electron-beam
  irradiated high-mobility metal-oxide-silicon transistors}.
\newblock Applied Physics Letters \textbf{110}, 123505 (2017).

\bibitem{Herman_1955}
F.~Herman.
\newblock \emph{The electronic energy band structure of silicon and germanium}.
\newblock Proceedings of the IRE \textbf{43}, 1703 (1955).

\bibitem{Ando_1982}
T.~Ando, A.~B. Fowler, and F.~Stern.
\newblock \emph{Electronic properties of two-dimensional systems}.
\newblock Reviews of Modern Physics \textbf{54}, 437 (1982).

\bibitem{Kruithof_1990}
G.~Kruithof and T.~Klapwijk.
\newblock \emph{Electron transport with two occupied subbands in a Si (100)
  inversion layer}.
\newblock Physical Review B \textbf{42}, 11412 (1990).

\bibitem{Nix1941thermal}
F.~Nix and D.~MacNair.
\newblock \emph{The thermal expansion of pure metals: copper, gold, aluminum,
  nickel, and iron}.
\newblock Physical Review \textbf{60}, 597 (1941).

\bibitem{Kawaji_1976}
S.~Kawaji, K.~Hatanaka, K.~Nakamura, and S.~Onga.
\newblock \emph{Mobility Hump and Inversion Layer Subbands in Si on Sapphire}.
\newblock Journal of the Physical Society of Japan \textbf{41}, 1073 (1976).

\bibitem{Eisele_1978}
I.~Eisele.
\newblock \emph{Stress and intersubband correlation in the silicon inversion
  layer}.
\newblock Surface Science \textbf{73}, 315 (1978).

\bibitem{Filatova_2019}
E.~O. Filatova, S.~S. Sakhonenkov, A.~S. Konashuk, S.~A. Kasatikov, and V.~V.
  Afanas’~ev.
\newblock \emph{Inhibition of Oxygen Scavenging by TiN at the TiN/SiO2
  Interface by Atomic-Layer-Deposited Al2O3 Protective Interlayer}.
\newblock The Journal of Physical Chemistry C \textbf{123}, 22335 (2019).

\bibitem{Bauer1972low}
G.~Bauer and H.~Kahlert.
\newblock \emph{Low-temperature non-ohmic galvanomagnetic effects in degenerate
  n-type InAs}.
\newblock Physical Review B \textbf{5}, 566 (1972).

\bibitem{Celik2011determination}
O.~Celik, E.~Tiras, S.~Ardali, S.~B. Lisesivdin, and E.~Ozbay.
\newblock \emph{Determination of the in-plane effective mass and quantum
  lifetime of 2D electrons in AlGaN/GaN based HEMTs}.
\newblock physica status solidi c \textbf{8}, 1625 (2011).

\bibitem{Sentaurus}
\emph{Sentaurus Process - Technology Computer Aided Design (TCAD) |
  Synopsys.”
  https://www.synopsys.com/silicon/tcad/process-simulation/sentaurus-process.html}.

\bibitem{Thorbeck_2015}
T.~Thorbeck and N.~M. Zimmerman.
\newblock \emph{Formation of strain-induced quantum dots in gated semiconductor
  nanostructures}.
\newblock AIP Advances \textbf{5}, 087107 (2015).

\end{thebibliography}

\begin{figure}
\centering
\includegraphics[scale=1]{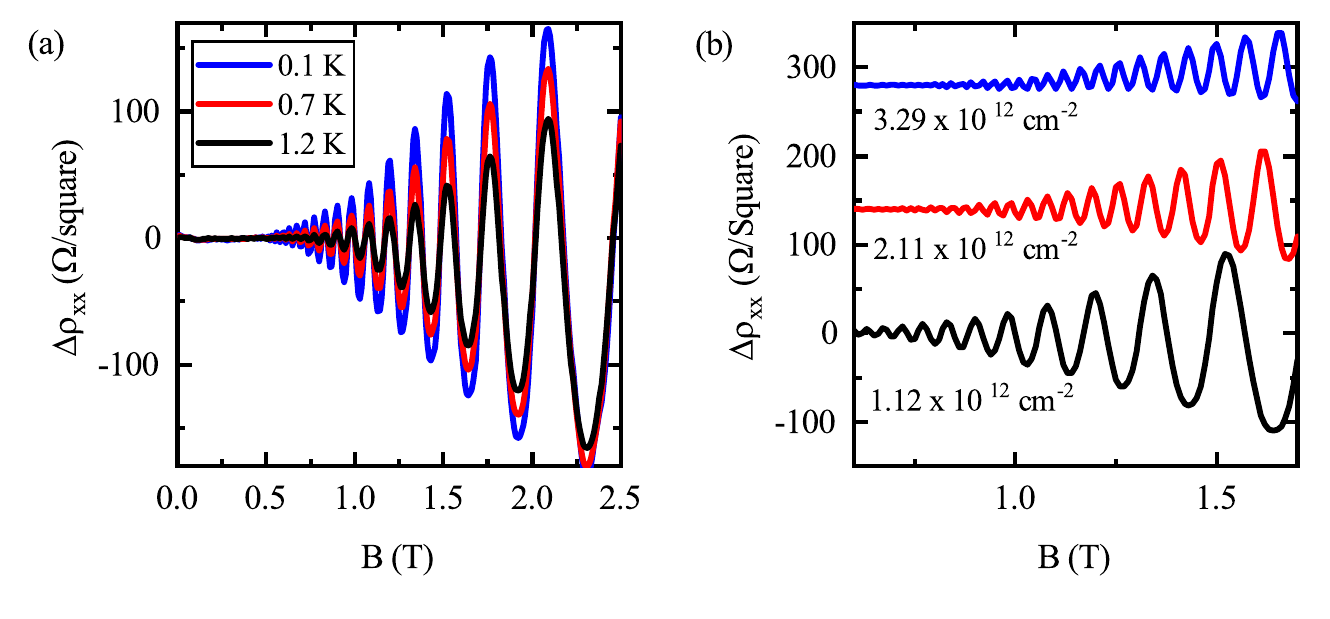}
\caption{\label{S_2}\textbf{SdH oscillations} (a) We can extract the effective mass by analysing the temperature dependent SdH oscillations at various magnetic field values $B$. (b) The envelope of the SdH oscillations in the fully spin and valley degenerate regime ($<$ \unit[2.5]{T}) we can extract the quantum lifetime $\tau _q$ for various densities. Curves are offset for clarity.}
\end{figure}

\begin{figure}
\centering
\includegraphics[scale=1]{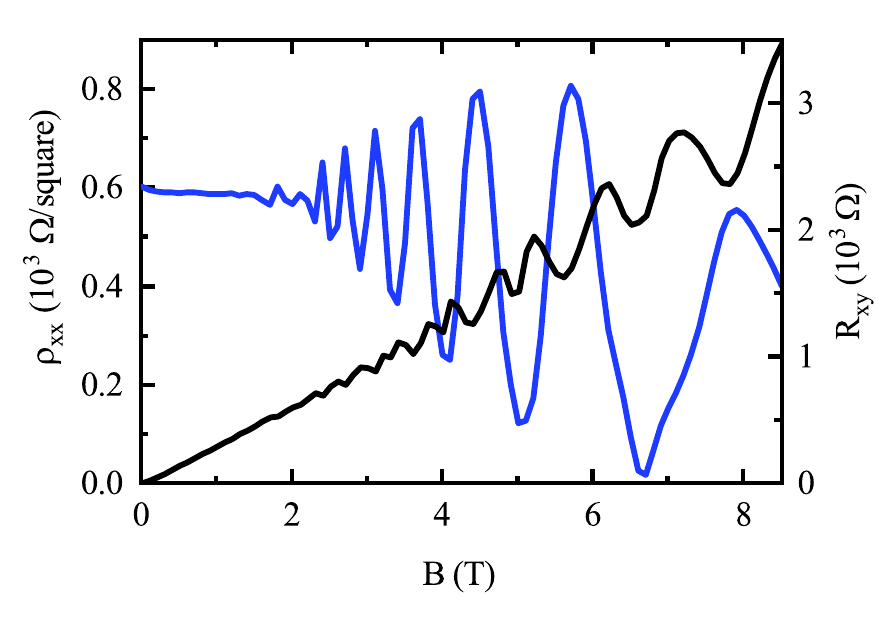}
\caption{\label{S_3}\textbf{TiN data at high fields} Longitudinal resistivity $\rho _{xx}$ (blue) and transverse Hall resistance $R_{xy}$ (black) of the sample measured with a TiN top gate at \unit[500]{mK}. Both the not well developed zeroes in $\rho _{xx}$ and the oscillating $R_ {xy}$ indicate multiple subband occupancy.}
\end{figure}

\end{document}